\documentclass[aps,prb,reprint,showpacs,
footinbib, superscriptaddress]{revtex4-1}

\usepackage{dcolumn}
\usepackage{epstopdf}
\usepackage{graphicx}
\usepackage{dcolumn}
\usepackage{bm}
\usepackage{bbm}
\usepackage{amsfonts}
\usepackage{amsmath}

\newcommand\redout{\bgroup\markoverwith{\textcolor{red}{\rule[.5ex]{2pt}{0.4pt}}}\ULon}

\usepackage[colorlinks=true,citecolor=blue]{hyperref}
\hypersetup{colorlinks=true,citecolor=blue,linkcolor=red,urlcolor=blue}

\begin{document}

\renewcommand{\thefigure}{\arabic{figure}}
\def\be{\begin{equation}}
\def\ee{\end{equation}}
\def\ber{\begin{eqnarray}}
\def\eer{\end{eqnarray}}

\def\kv{{\bf k}}
\def\qv{{\bf q}}
\def\pv{{\bf p}}

\def\sigmav{{\bf \sigma}}
\def\tauv{{\bf \tau}}

\newcommand{\h}[1]{{\hat {#1}}}
\newcommand{\hdg}[1]{{\hat {#1}^\dagger}}
\newcommand{\bra}[1]{\left\langle{#1}\right|}
\newcommand{\ket}[1]{\left|{#1}\right\rangle}

\title{ Superconducting electron and hole lenses}
\date{\today}

\author{H. Cheraghchi}
\affiliation{School of Physics, Damghan University, 36716-41167, Damghan, Iran}

\author{H. Esmailzadeh}
\author{A. G. Moghaddam}\email{agorbanz@iasbs.ac.ir}
\affiliation{Department of physics, Institute for Advanced Studies in Basic Sciences (IASBS), Zanjan 45137-66731, Iran}

\begin{abstract}
We show how a superconducting region (S) sandwiched between two normal leads (N), in the presence of barriers, can act as a lens for propagating electron and hole waves by virtue of the so-called crossed Andreev reflection (CAR). The CAR process which is equivalent to the Cooper pair splitting into the two N electrodes provides a unique possibility of constructing entangled electrons in solid state systems. When electrons are locally injected from an N lead, due to the CAR and normal reflection of quasiparticles by the insulating barriers at the interfaces, sequences of electron and hole focuses are established
inside another N electrode. This behavior originates from the change of momentum during electron-hole conversion beside the successive normal reflections of electrons and holes due to the barriers. The focusing phenomena studied here is fundamentally different from the electron focusing in other systems like graphene \emph{pn} junctions. In particular due to the electron-hole symmetry of superconducting state, the focusing of electrons and holes are robust against thermal excitations. Furthermore the effect of superconducting layer width, the injection point position, and barriers strength is investigated on the focusing behavior of the junction. Very intriguingly, it is shown that by varying the barriers strength, one can separately control the density of electrons or holes at the focuses.
\end{abstract}


\maketitle

\section{Introduction}
One of the fundamental aspects of quantum mechanics is non-locality which manifests as non-classical correlations between spatially separated particles which is usually called as quantum entanglement \cite{horodecki}. During the last decades a tremendous amount of efforts has concentrated to exploit the entanglement as the main resource for quantum information processing and its technological applications \cite{nielsen}. In solid state systems among variety of proposed methods to create entanglement between electrons, one of the most accessible sources of entangled electrons is  Cooper pair (CP) splitting \cite{blatter01,recher01}. In conventional superconductors CPs are composed of two electrons in a spin singlet state which can be used to produce two entangled electrons. The process of spatially separating two electrons of a CP without losing their correlations is commonly called as crossed Andreev reflection (CAR) which takes place in the system of a superconductor (S) coupled to two normal leads (N) \cite{flatte95,deutscher00,deutscher02,hekking01, feinberg2003, zaikin2007}. In this process an incoming electron from one N lead is able to enter to the S region via attracting another electron from the other N lead which leaves a hole inside it.
\par
Unlike Andreev reflection (AR) which takes place locally at any NS interface, CAR has been proved to be hardly detectable in experiments, due to the cancellation of its contribution with that of the elastic co-tunneling (EC) in the cross conductance measurements \cite{morpurgo05,zimansky06,zimansky09,schonenberger09epl}. Among proposed setups to detect CAR, one method which has been experimentally verified, uses two quantum dots between each N lead and the superconductor. \cite{schonenberger09,stunck10,schonenberger11,schonenberger12,das12}. In other proposals, the replacement of normal leads with strong ferromagnets (F) or even spin half metal has been suggested \cite{maekawa03,beckmann04,cadden07}. Alternatively, some other works have focused on the diffusive regime of transport in NSN structures \cite{belzig06,golubov06,zaikin09}. Other groups have revealed that the Coulomb interactions, in particular in the quantum dot systems, can lead to the nonlocal Andreev processes \cite{konig09,wysokinski13}. Moreover, very recently it has been suggested to use devices based on graphene or topological insulators and generally the two dimensional materials in which the low energy excitations behave as Dirac fermions \cite{cayssol08,nilsson08,linder09,waintal10,xing11,linder14,recher15}.
\begin{figure}[tb]
\includegraphics[width=0.99\linewidth]{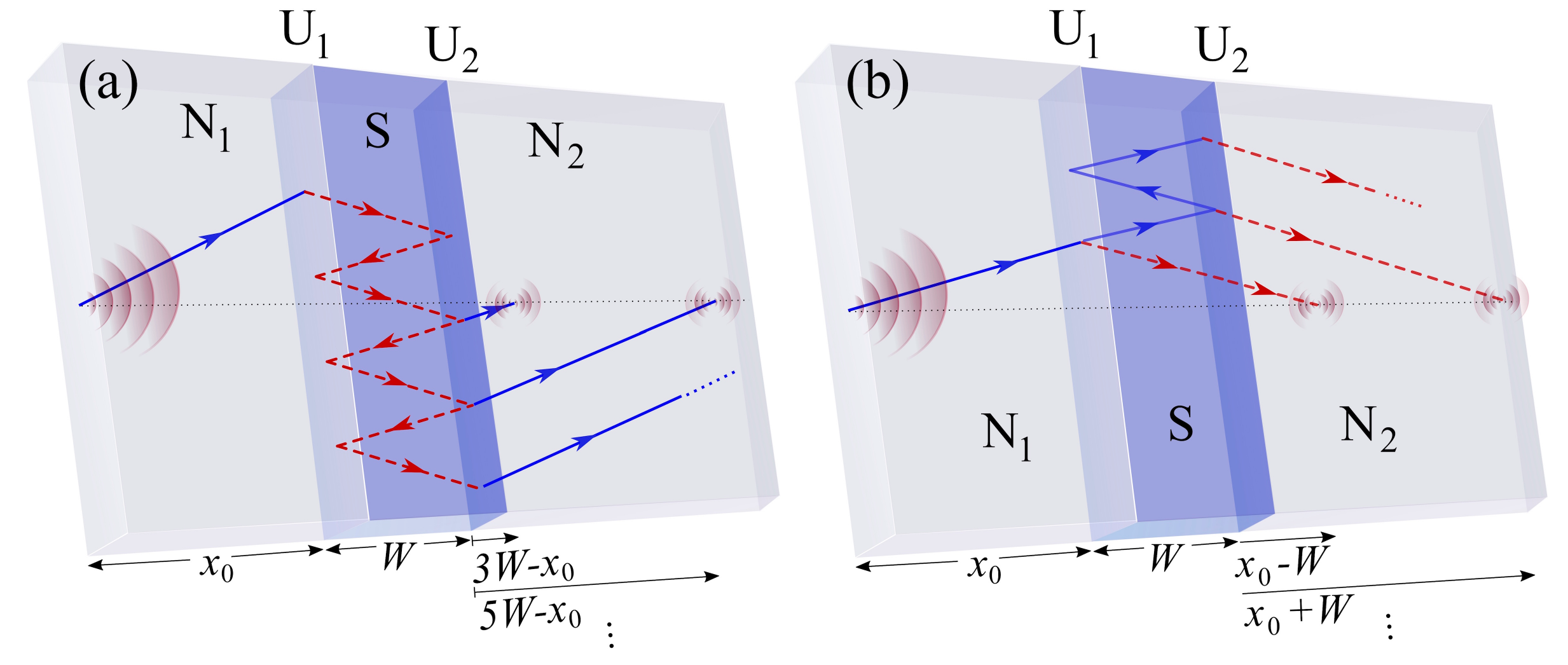}
\caption{(Color online) A schematic view of the superconducting lens is shown beside the classical trajectories of electrons (solid blue lines) and holes (dashed red lines) originating from a point source inside N$_1$. The focusing effect is caused by the crossed Andreev reflection (nonlocal electron-hole conversion) in combination with normal reflections which take place at the interfaces. It must be noticed that for clarity only the trajectories with certain angle are shown. However according to the conservation rules and simple geometric relations, the positions of focal points where the trajectories cross the $x$ axis do not change by varying the angle.}
\label{fig1}
\end{figure}
\par
In this paper we investigate the focusing effect for both electrons and holes injected locally from an electron source and passing through a planar normal-insulator-superconductor-insulator-normal (NISIN) junction. Transport properties of such ballistic systems and particularly the effect of barriers on CAR and EC signals have been already investigated \cite{hekking01, feinberg2003, zaikin2007}. Here we will show any planar ballistic NISIN junction under certain circumstances can lead to the focusing of both electrons and holes. The underlying mechanism for the focusing property here is the combination of CAR or equivalently non-local electron-hole conversion (EHC) induced by superconducting correlations with large momentum transfer due to the potential barriers at the interfaces. The main characteristic of NISIN junctions is depicted schematically in Fig \ref{fig1} where some semiclassical trajectories of electrons and holes originating from a point source are shown.  
It clearly shows that electrons and holes trajectories are completely separated since the momentum and velocity directions for electron and hole excitations are parallel and antiparallel, respectively. When electronic waves injected from a point source inside N$_1$ arrive at an NS interface, they are partially reflected and transmitted as both electrons and holes where the last scattering process occurs due to  EHC. In other words there exist four different scattering processes in  NISIN junctions including normal and Andreev reflections which take place locally, as well as EC and CAR which correspond to the transmission as an electron and a hole, respectively. In particular for the transmitted waves each EHC process results in the negative refraction of electronic wave function and then successive scattering processes can give rise to the sequence of electron and hole focal points inside N$_2$ (see Fig. \ref{fig1}). 
\par
It must be mentioned that prior to the present work, G\'omez et al. have shown that an NSN junction based on graphene can act as a lens for electron and hole excitations \cite{yeyati12}. However in their investigation the lensing property is mostly accompanied by the chiral relativistic characteristics of the charge carriers and the so-called Klein tunneling which occurs at the \emph{pn} type interfaces in graphene. In fact even without the superconducting correlations, the Dirac systems like graphene has been proven to show negative refraction at the \emph{pn} interfaces and they can behave as 
Veselago-like lenses \cite{cheianov07,moghaddam10}. On the other hand, here we show that in NISIN junctions, the combination of EHC with the effect of barriers can lead to the electron and hole focusing. This phenomenon does not require a special Dirac band structure of graphene or topological insulators and it can be in principle observed in the conventional superconducting devices as well as those based on the Dirac materials.
\par
The paper is organized as follows. After the introduction, in Sec. \ref{sec2}, the model consisting of superconductor between normal metals and at the presence of barriers is introduced. The formulation which is based on the Gorkov's equations and Green's functions are presented afterwards. In Sec. \ref{sec3} we show the numerical results for local particle and current densities. These results reveal the electron and hole focusing phenomena in the NISIN structure. Our findings are followed by the discussion over the basic underlying mechanisms, effect of various parameters and irregularities on the results, and possible experimental realizations. Finally, Sec. \ref{sec4} is devoted to the conclusions.

\section{Theoretical model and formalism}
\label{sec2}
In order to have a quantitative description of the anticipated focusing effects in NISIN junctions, we exploit the Green's function formalism. In superconducting structures and for the time-independent situations, the full Green's function $\hat{\cal G}({\bf r},t|{\bf r'},t')=\int d\omega \hat{\cal G}_{\omega}({\bf r}|{\bf r'})\exp[-i\omega(t-t')]$ is defined with the following matrix form in the Nambu space and the energy representation, 
\begin{eqnarray}
\hat{\cal G}_{\omega}({\bf r}|{\bf r'})=
\begin{pmatrix}
\ g({\bf r}|{\bf r}' ;\omega) & f({\bf r}|{\bf r}' ;\omega)\\ f^{\dagger}({\bf r}|{\bf r'};\omega) & -g({\bf r}'|{\bf r};\omega)\end{pmatrix},
\end{eqnarray}
in which 
\begin{eqnarray}
g({\bf r}|{\bf r'};\omega)=\int d\tau e^{i\omega \tau}\langle \hat{\psi}_s^\dag ({\bf r},\tau) \hat{\psi}_s ({\bf r'},0) \rangle,\\
f({\bf r}|{\bf r'};\omega) =-
\int d\tau e^{i\omega \tau}
\langle \hat{\psi}_\uparrow ({\bf r},\tau) \hat{\psi}_\downarrow ({\bf r'},0) \rangle,  
\end{eqnarray}
are  the so-called normal and  anomalous parts of the Green's function. Here field operators $\hat{\psi}_s^\dag ({\bf r},\tau)$ and $\hat{\psi}_s ({\bf r},\tau)$ represent the creation and annihilation of excitations with spin $s$ at point ${\bf r}$ and time $\tau$. The statistical averaging is denoted by $\langle\cdots\rangle$. In ballistic regime, the full Green's function can be obtained from the matrix form of Gorkov equations \cite{fetter},
\begin{eqnarray}\label{gorkov}
[\omega   -{\cal H}({\bf r}) \hat{\tau}_z  +\Delta({\bf r})\hat{\tau}_x -\hat{\cal V}_{\rm ext} ]
\hat{\cal G}_{\omega}(\textbf{r}|\textbf{r}^\prime)=  \delta(\textbf{r}-\textbf{r}^\prime),
\end{eqnarray}
with Pauli matrices $\hat{\tau}_z $ and $\hat{\tau}_x$ operating over the Nambu (electron-hole) space.
The single particle Hamiltonian ${\cal H}({\bf r}) $ describes a typical metal with quadratic dispersion relation as, 
\begin{eqnarray}
{\cal H} ({\bf r})=-\frac{\hbar^2}{2m} \nabla^2  + {\cal U}({\bf r}) -\mu,
\end{eqnarray}
in which $\mu$ indicates the Fermi energy. For the superconducting gap we use the sharp interface model in which the gap has a constant finite value $\Delta({\bf r})=\Delta_0$ inside the superconductor ($0<x<W$) and vanishes outside it. In fact the more precise spatial form of the pair potential should be obtained using the self-consistency relation of the gap $\Delta({\bf r})=-\lambda \int{d\omega} f({\bf r}|{\bf r};\omega)$. However due to the presence of barriers at the interfaces, the superconducting gap inside S is only weakly affected by the adjacent N regions and therefore using self-consistency equation to obtain the spatial variations of $\Delta({\bf r})$ is not crucial.
The potential function $ {\cal U}({\bf r})=[U_1\delta(x-0)+U_{2}\delta(x-W)]W$ accounts for the two barriers at the interfaces induced by atomically thin insulating layers. In the absence of external potential which represents the electron injection, the Gorkov equation can be solved to obtain the unperturbed Green's function $\hat{\cal G}_0({\bf r}|{\bf r'})$. Then in the presence of external perturbation $\hat{\cal V}_{\rm ext}(V_0/2) (\hat{\mathbbm 1}+\hat{\tau}_z) \delta({\bf r}-{\bf r}_0)$,
the full Green's function $\hat{\cal G}({\bf r}|{\bf r'})$ can be obtained using the Dyson equation,  
\begin{eqnarray}\label{dyson}
\hat{\cal G}({\bf r}|{\bf r'})= \hat{\cal G}_0({\bf r}|{\bf r'})+\int d{\bf x}\,\hat{\cal G}_{0}({\bf r}|{\bf x}) \hat{\cal V}_{\rm ext}({\bf x})   \hat{\cal G}({\bf x}|{\bf r'}).
\end{eqnarray}
\par
From the Green's function one can calculate all of the local properties of the system, including local densities and currents in the framework of linear response theory.
In particular, the electron and hole local density of states (LDOS) can be extracted from the diagonal components of matrix $\hat{\cal G}_{\omega}  ({\bf r}|{\bf r})$, namely using the relations $
n_e({\bf r};\omega)={\rm Im}  {\cal G}^{ee}_{\omega}({\bf r}|{\bf r})$ and $
n_h({\bf r};\omega)={\rm Im}  {\cal G}^{hh}_{\omega}({\bf r}|{\bf r})$. Then assuming weak electron injection from the source at ${\bf r}_0$ and expanding over $V_0$, the lowest order corrections of the electrons and holes LDOS read,
\begin{eqnarray}\label{ehd}
\delta\,n_{e}({\bf r};\omega) \propto {\rm Im} [g_0({\bf r}|{\bf r}_0 ;\omega)g_0({\bf r}_0|{\bf r} ;\omega)],\\
\delta\,n_{h}({\bf r};\omega)\propto {\rm Im} [f_{0}^{\dag}({\bf r}|{\bf r}_0 ;\omega)f_0({\bf r}_0|{\bf r};\omega)],
\end{eqnarray}
where $g_0$ and $f_0$ indicate the normal and anomalous parts of the unperturbed Green's function. On the same ground one can immediately see that the magnitudes of local current densities, up to linear order in the injected current at the source point, are given by, 
\begin{eqnarray}\label{currents}
J_{e}({\bf r};\omega)\propto| g_0({\bf r}|{\bf r}_0 ;\omega)g_0({\bf r}_0|{\bf r} ;\omega) |, 
\\
J_{h}({\bf r};\omega)\propto| f_{0}^{\dag}({\bf r}|{\bf r}_0 ;\omega)f_0({\bf r}_0|{\bf r};\omega)|.
 \end{eqnarray}
\section{Numerical results and discussion}
\label{sec3}
In the following we will present the numerical results obtained by discretized Green's function calculation for LDOS and current densities. 
Using the translational symmetry along vertical $y$ direction, the unperturbed Green's function can be written in the Fourier transformed form,
\begin{eqnarray}\label{fourier}
\hat{\cal G}_0({\bf r}|{\bf r'})=\int \frac{dq}{2\pi} e^{-iq(y-y')} \hat{\cal G}_0(x|x';q).
\end{eqnarray}
Hence, Eq. (\ref{gorkov}) reduces to a one dimensional form for $\hat{\cal G}_0(x|x';q)$ which can be solved in general by discretization methods. The only important point is to set the increment $\delta x$ small enough in comparison with the Fermi wavelength $\lambda_F$ which is the smallest length scale of the problem. Finally by numerical implementation of the two integrations in Eqs. (\ref{fourier}) and (\ref{dyson}) the perturbed Green's function at the presence of the electron source can be obtained.
\par
In order to have observable CAR signal, the superconductor must have width $W$ comparable with the superconducting coherence length $\xi=\hbar v_F/\Delta_0$ in which $v_F$ is the Fermi velocity \cite{zaikin2007}. In fact the nonlocal EHC is efficient only when $W\sim \xi$ which increases the probability of CAR. For thinner or very wide superconducting layers the CAR is weakened in favor of EC and local AR, respectively.
In typical superconductors with $\Delta_0\lesssim 1$meV, the superconducting correlations extend over $\xi\sim 0.1 -1 \mu$m, which gives the typical size of the proposed superconducting lenses. In addition, the presence of potential barriers at the interfaces $x=0,W$ leads to the further suppression of the local AR and EC through the junction and subsequently increases the efficiency of electron or hole waves focusing mediated by the combination of CAR and normal reflections from the barriers. 
\par
In the following the electron and hole focusing property of  a superconducting NISIN heterostructure is examined by studying the spatial variations of the induced local density and local currents inside the second N region. We further investigate the  influence of width of superconducting region, the distance of the electron injection from the interfaces, and the strength of the barriers at the interfaces on the focusing properties of the superconducting lens. Finally we comment on the experimental feasibility of the theoretical model and discuss over the obtained results and their limitations.
\begin{figure}[tb]
~~~\includegraphics[width=0.99\linewidth]{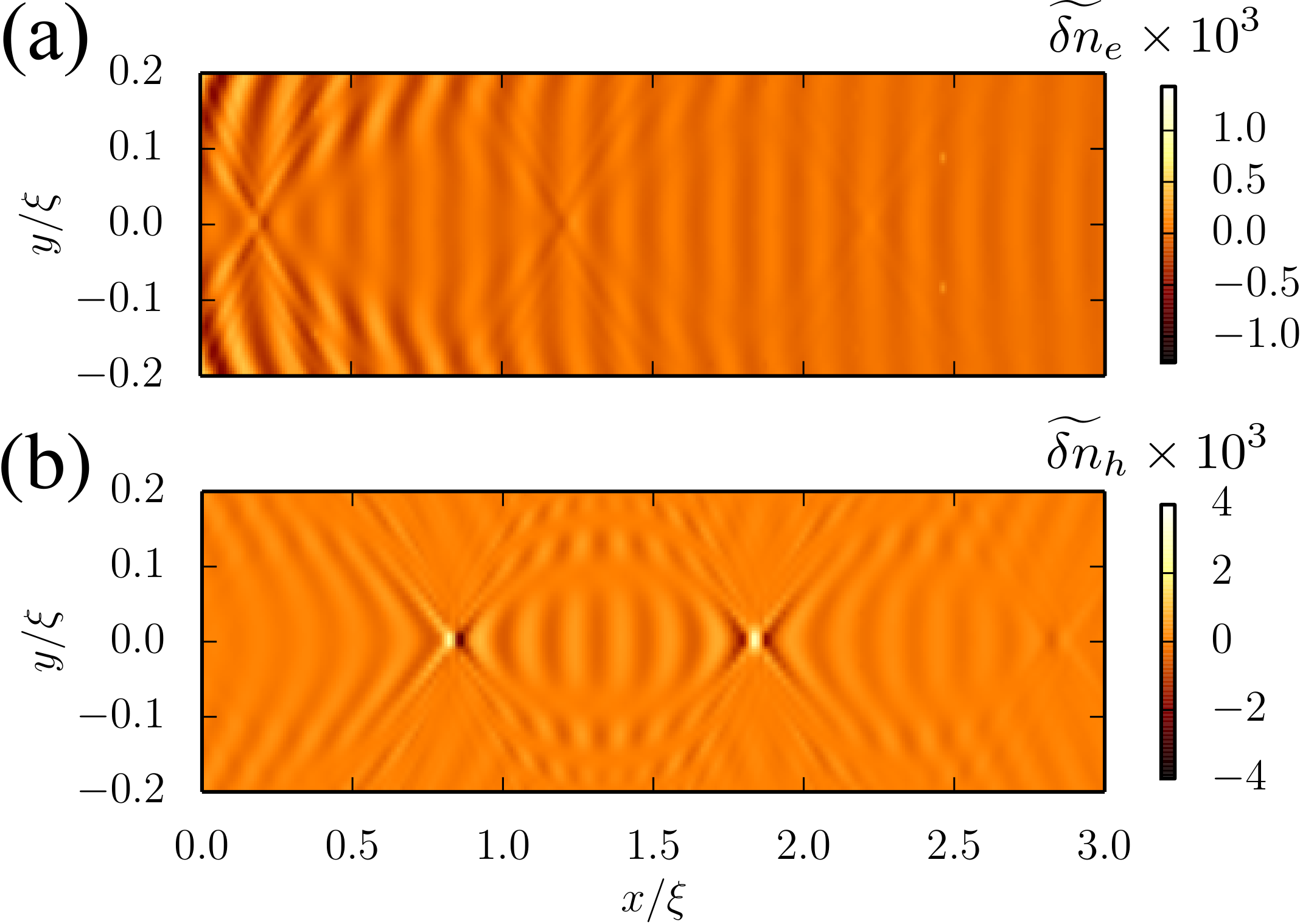}
\caption{(Color online) Spatial variations of (a) electron (b) hole scaled excess densities defined by $\widetilde{\delta n}=\delta n/(V_0 n_0^2)$. The electron injector is located at $x_{0}=-2.7W$ inside the first N lead, measured from the first interface. The density profiles are plotted inside the second normal contact N$_2$. The superconducting layer has a width $W=0.5\xi$ and the potential barriers at the interfaces are $U_1=U_2=2 \mu$ and other parameters are $\Delta_0/\mu=.01$, $\epsilon/\Delta_0=0.8$. For both electrons and holes sequences focal points along the $x$-axis can be observed. The electron density shows an extra decaying push which indicates the EC contributions. The oscillations in the densities are due to the quantum interference of electronic waves upon scattering from the interfaces and the superconducting layer.}
\label{fig2}
\end{figure}
\subsection{Electron and hole focusing}
In order to see the focusing signature of the superconducting layer, the spatial variations of the LDOS for the electrons and holes are shown in Fig. \ref{fig2}. 
We define the dimensionless excess density of states denoted by $\widetilde{\delta n}({\bf r})=\delta n ({\bf r})/ (V_0 n_0^2)$ in which $n_0=m/(\pi\hbar ^2)$ is the constant density of states of the two dimensional electron gas (2DEG).
The width of S layer is assumed to be $W=0.5\xi$ and the source is placed at $x_0=-2.7 W$. Now the sequences of the focal points where electrons and holes are focused, can be clearly seen in Fig. \ref{fig2}. These focal points are accompanied by oscillations around them which originate from the quantum interference of refracted electronic waves. In the case of electron density due to the EC, there is an extra overall push function which shows oscillatory decreasing dependence with the distance from the second interface. On the other hand, the focuses followed by oscillations around them can be seen in the hole density. As we will discuss in the following, the focuses of either electrons or holes originate from the combination of EHC at the N/S interfaces and normal reflection of bogoliubov quasiparticles inside the superconductor. It must be emphasized that the second process is only possible when the barriers are present. Moreover without any barrier at the interfaces the CAR itself will be suppressed in favor of the AR and EC processes. 
\par
It is instructive to describe the appearance of electron and hole focuses based on a semiclassical argument. As illustrated in Fig. \ref{fig1}, the trajectories of propagating electronic waves, originated from the injection point $x_0$ inside N$_1$, are refracted at the N/S interfaces when an EHC takes place. It is well-known that the velocity of a hole as an excitations below the Fermi level have opposite direction with respect to its momentum. However the velocity and momentum of the electrons have the same direction. Therefore when an EHC occurs, because of momentum conservation along $y$ direction, the electronic waves undergo a negative refraction which is revealed in the classical trajectories as well. 
The negative refraction here is a reminiscent of the famous Veselago lenses in optics which are made at the interfaces between a conventional dielectric medium and a metamaterial with negative refraction index. Now the refractions caused by EHC and successive normal reflections of the Bogoliubov quasiparticles from the barriers, can result in a sequence of focuses inside the lead $N_2$. It should be noticed that the presence of barriers are not only crucial for the normal reflections but also they provide the source of the large momentum transfer in the EHC processes.
In Fig. \ref{fig1}, for the sake of clarity, only those trajectories
which lead to a focal point are shown. 
However there are other trajectories corresponding to the normal and Andreev reflections back into N$_1$ as well as the diverging waves of electrons and holes inside N$_2$ which are not shown here. 
\begin{figure}[tp]
\includegraphics[width=0.95\linewidth]{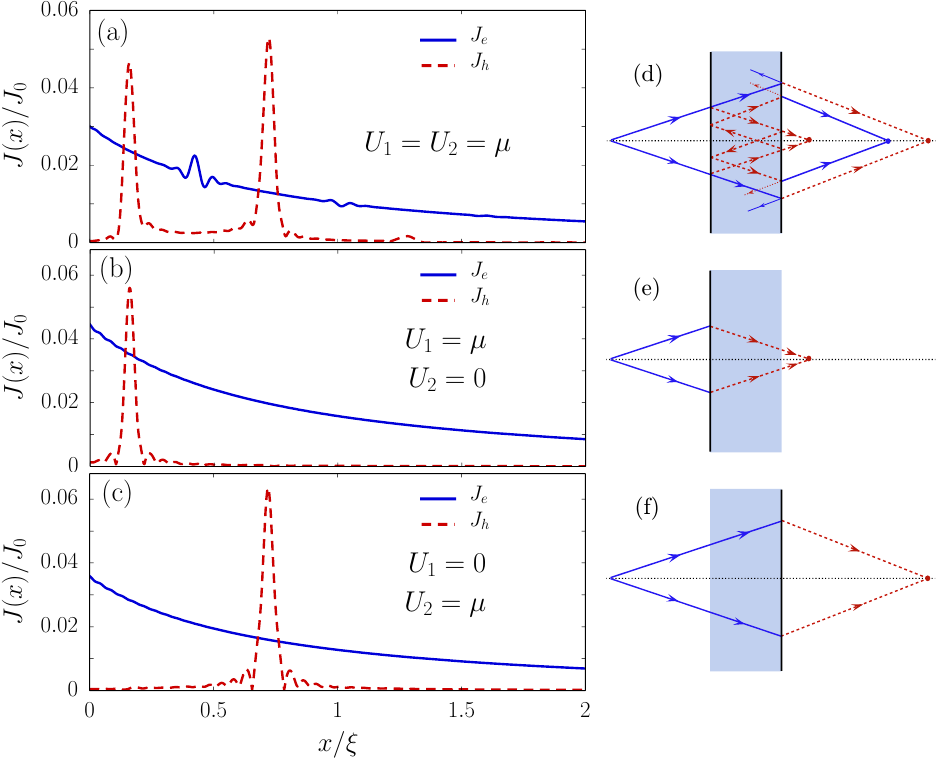}
\caption{(Color online) 
Local current densities of electrons ($J_e$) and holes ($J_h$) inside N$_2$ along the $x$ axis and versus the distance from the second interface when (a) $U_1=U_2=\mu$, (b) $U_1=\mu$, $U_2=0$, and (c) $U_1=0$ and $U_2=\mu$.
(d)-(f) indicate the corresponding schematics in which the classical trajectories leading to the electron and hole focuses are shown. The electron and hole propagations are indicated by solid blue and dashed red lines, respectively.
The width of the superconductor and injection point position are $W/\xi=0.22$ and $x_0/\xi=0.35$. The superconducting gap and excitation energy are assumed as $\Delta_0/\mu=0.01$ and $\epsilon/\Delta_0=0.8$.
This figure clearly indicates that the sequence of focuses can be observed when both interfaces have a limited transparency ($U\neq 0$). However when the second or first interfaces are fully transparent ($U=0$) only first and second hole focuses survive, respectively.
}
\label{fig3}
\end{figure}
\par
The semiclassical trajectories of electrons and holes can be helpful in finding the positions of the focal points. It must be mentioned that although the trajectories are defined by both the source point and the angle of incidence, however the positions of the focuses do not vary by the angle of trajectories, warranted by the simple trigonometric relations. 
Depending on the number of EHCs during a process (being even or odd), an electron or a hole focus shows up which can be seen from Fig. \ref{fig1}(a) and (b), respectively. When the number of normal reflections back into the superconductor increases, new focuses far from the second interface can be established. Then, in principle, we end up with two infinite sequences of separate electron and hole focuses. Using the trigonometric relations the positions of the electron and hole focuses are obtained as below,
\begin{equation}
\label{focuses}
\begin{array}{l}
x_{e}^{(n)}=(2n+1)W-x_{0} \\  
x_{h}^{(n)}=(2n-1)W+x_{0}   
\end{array}~~;~~n=0,1,2,...
\end{equation}
One should note that the first electron focus corresponds to $n_{e;1}=\lfloor(x_0/2W)+0.5\rfloor$ while the hole focuses start from $n_{h;1}=1$ or $0$ depending on the width $W$ being larger or smaller than $x_0$, respectively.
\begin{figure}[tp]
\includegraphics[width=0.8\linewidth]{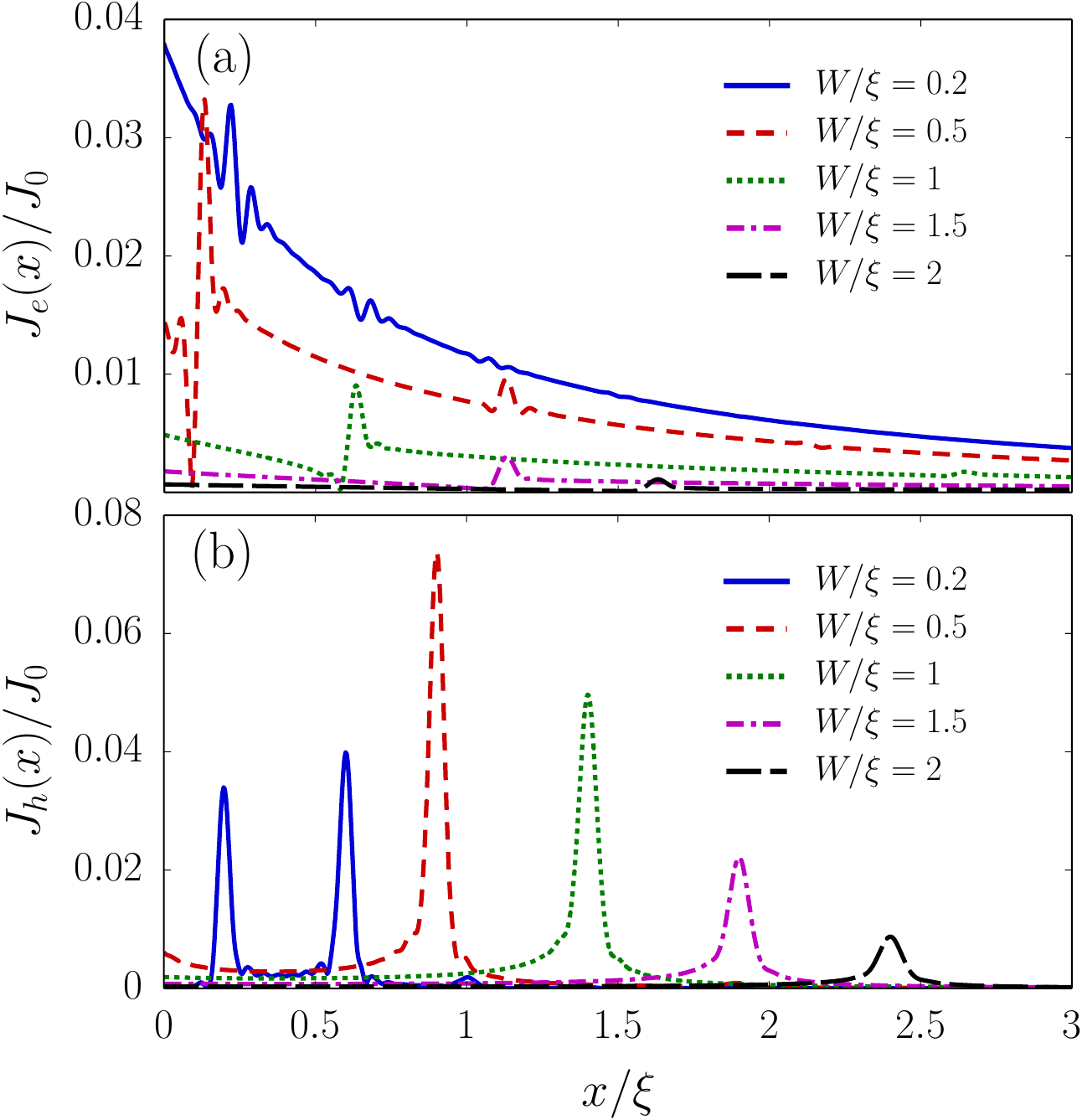}
\caption{(Color online) Spatial variation of the local current densities for (a) electrons and (b) holes versus the distance from the second interface along the $x$ axis for various values of the superconducting layer width $W$. Although the EC signal indicated by electron current $J_e$ shows an overall decline for wither $W$, the CAR signal given by hole current $J_h$ is strong for $W/\xi\lesssim 1$. The parameter used for this plot have the values 
$x_0/\xi=0.4$, $U_1=U_2=\mu$, $\lambda_F/\xi=0.04$, and $\epsilon/\Delta_0=0.8$.} \label{fig4}
\end{figure}

\subsection{Electron and hole local current densities}
Very similar to the LDOS, the local current densities of electrons and holes denoted by $J_e$ and $J_h$ can
indicate the focusing signature of the superconducting layer between two normal electrodes. The intensities of electron and hole currents with the corresponding focal points are shown in Fig. \ref{fig3}. There are also small oscillations around the focuses but not as clear as the Friedel type oscillations revealed in the LDOS spatial variations. Here a single geometry but for different configurations of the barriers is considered, in order to see the effect of barriers more clearly. Figures \ref{fig3} (a)-(c) correspond to the cases with two barriers, single barrier at the first interface, and finally a single barrier at the second interface with corresponding sketches for the electron and hole trajectories. We see that in the presence of both barriers the first two focuses of the holes are very strong. These two focal points corresponds respectively, to the EHC at the first and second interfaces. 
Other focuses for the holes as well as those of electrons are much weaker since they originate from higher order processes when successive EHC and normal reflections take place.
When one of the barriers are absent then the normal reflections cannot lead to further focuses. Subsequently the EHC is sufficiently strong only at one of the interfaces which leads to a single hole focus without any electron focusing. This suggest a way to control the intensity of different focuses by varying the strength of the two barriers, separately.
\par
As mentioned earlier, the CAR is very effective for S layers with certain range of the widths being on the order of the coherence length ($W\sim \xi$). Hence, we expect that varying the width of S layer, can strongly influence the focusing properties. This can be seen clearly in Fig. \ref{fig4} where the spatial variations of the electron and hole local currents for various values of $W/\xi$ are shown. Starting from a thin S layer with $W/\xi=0.2$,
EC processes is dominant and therefore the overall push of the electron current is larger as indicated in Fig. \ref{fig4}(a). By increasing the width the EC signal weakens and due to the dominance of CAR the electron and hole focuses are more intense for the junctions with intermediate widths e.g. $W/\xi=0.5$. Then further increase in the width, weakens the CAR in favor of local AR and subsequently the intensity of the currents and excess densities around the focuses will be suppressed. For very long junctions $W\gg \xi$ only normal reflection and AR could take place and CAR as well as the focusing signature diminish. As another interesting point it must be mentioned that when the width of the junction is smaller than the distance of the injection point from the first interface ($x_0$), two first hole focuses are very intense. These focuses originate from the EHC at the first and second interfaces, respectively and their positions are $x^{(0)}_h=x_0-W$ and $x^{(1)}_h=x_0+W$
according to Eq. (\ref{focuses}). On the other hand, when the width is larger than the distance $x_0$, the EHC at the first interface cannot give rise to a hole focus inside N$_2$, as can be understood from Fig. \ref{fig1}(b).
Therefore only one single hole focus located at $x_h^{(1)}$ becomes intense when $W>|x_0|$. Other hole focuses with $n_h\geq 2$ as well as those of electrons which are related to the higher order processes have much weaker intensities as explained before.
\begin{figure}[tp]
\includegraphics[width=0.8\linewidth]{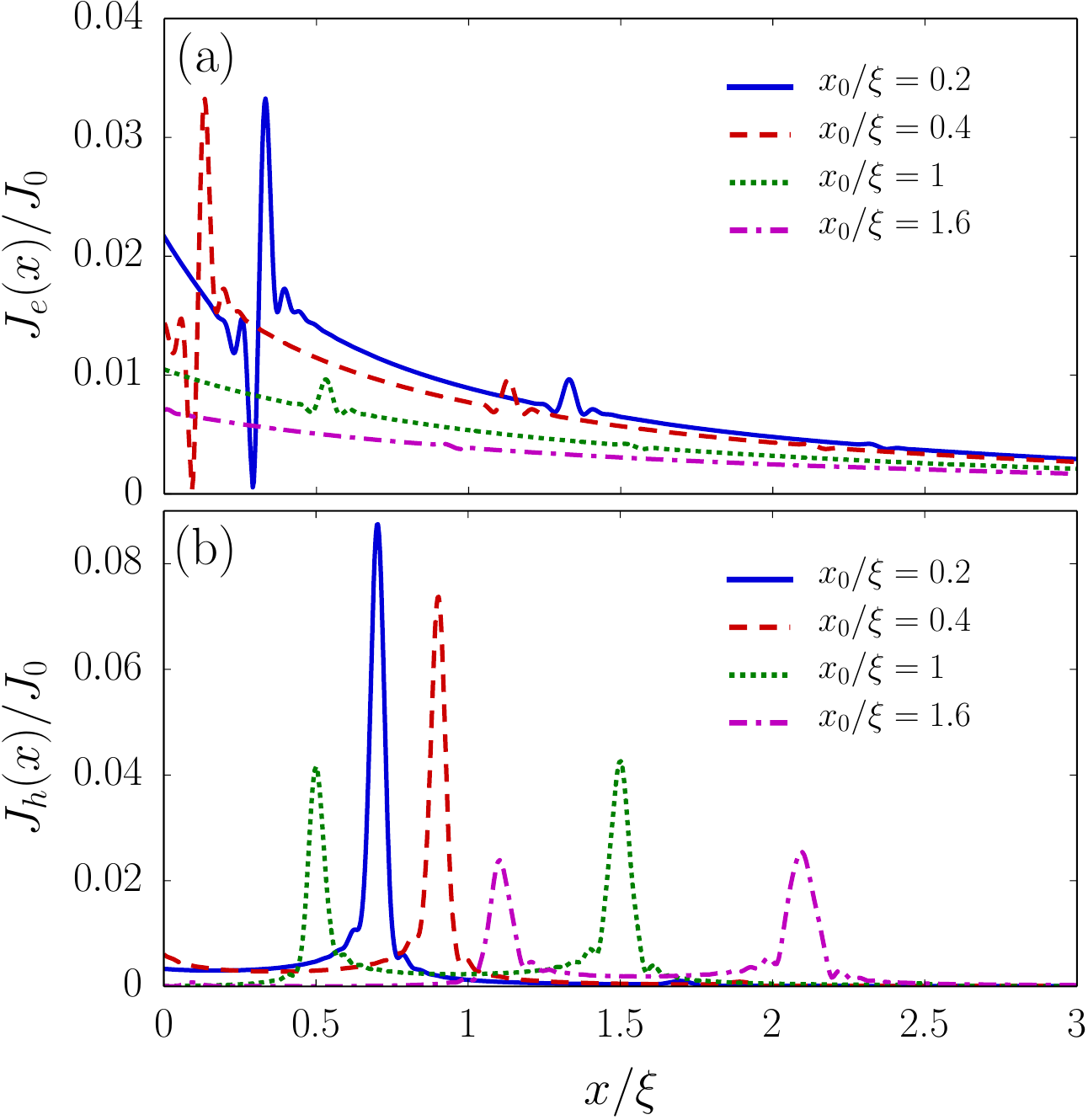}
\caption{(Color online) Spatial variation of (a) local electron current and (b) hole current along the $x$ axis for various distances of the injection point from the first interface ($x_0$). The width of the superconductor is $W/\xi=0.5$ and other parameters are the same as those in Fig. \ref{fig4}.}
\label{fig5}
\end{figure}
\par
To see the effect of the injection point distance on the EC and CAR signals, Fig. \ref{fig5} shows the spatial variations of  $J_e$ and $J_h$ for some different values of $x_0/\xi$ when the width is $W=0.5\xi$. 
The results show that both electron and hole currents with main contributions from EC and CAR decreases if the injection point becomes farther away from the S layer. Moreover as it has been mentioned above, depending on the distance of the injection point being smaller or larger than $W$, there will be a single or two intense hole focuses, respectively. On the same ground, when the injection point is close to the S layer ($|x_0|<W$), the first electron focus which originates
from two successive EHC at the two interfaces is very strong. However in the opposite limit of far injection points ($x_0>W$), beside EHC processes, two or more reflections of Bogoliubov excitations from the interfaces back into S layer take place to establish the first electron focusing. Therefore the intensity of resulting electron focuses will be much weaker as it is clear from Fig. \ref{fig5}(a). 
\par
Before closing this part, we should comment on the effect of the excitation energy $\epsilon$ on the focusing properties of the NISIN junctions. Throughout the paper the results are presented for the certain value of energy inside the superconducting gap, $\epsilon=0.8\Delta_0$. However we have checked the dependence of the results on $\epsilon$ and it has been found that the change in the excess densities $\delta n_{e,h}$ and currents $J_{e,h}$ due to the  excitation energy varied from $0$ to $\Delta_0$ is very small. This is a simple consequence of the fact that the superconducting gap and therefore $\epsilon$ is very smaller than the Fermi energy $\mu$ and all of the wave vectors $k_{e,h}=\sqrt{(2m/\hbar^2)(\mu\pm\epsilon)}$ can be approximated with $k_F=\sqrt{(2m/\hbar^2)\mu}$. So the scattering and focusing properties of the junction are determined only by the geometrical parameters $W$ and $x_0$ as well as the Fermi wave vector, almost irrespective of the excitation energy. Above the superconducting gap ($\epsilon>\Delta_0$) the EHC is suppressed and subsequently the focusing patterns disappear when $\epsilon\gg\Delta_0$.
\subsection{The effect of barriers strength}
In what follows, we examine the influence of barriers strength in more details. As it has been revealed by our results in the previous part, the presence of at least one potential barrier at the interfaces is crucial for the emergence of hole focusing. On the other hand the electron focusing will be absent unless we have effective barriers at both interfaces. 
In the absence of potential barriers, the dominant scattering processes are AR and EC where both of them satisfy the momentum conservation. The presence of barriers at the interfaces favors the normal reflection and CAR which involve a finite momentum transfer and simultaneously suppresses the other scattering processes i.e. local AR and EC.
\begin{figure}[tp]
\includegraphics[width=0.8\linewidth]{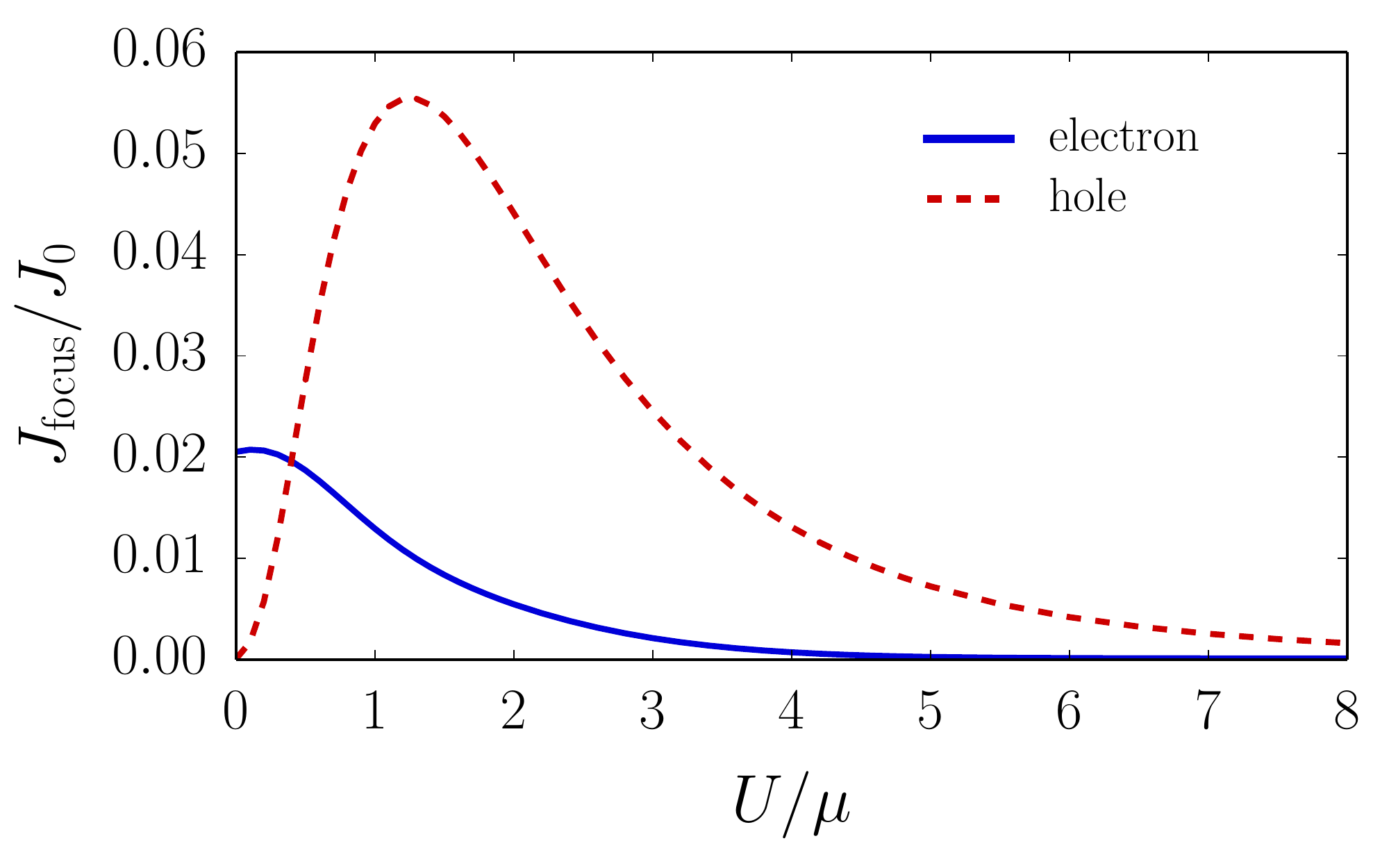}
\caption{(Color online) Dependence of the hole and electron current densities at the first hole focus ($x_h^{(0)}$) on the barriers strength $U$. Although the electron current decreases monotonically with $U$ the hole current at the focus becomes maximum at a certain strength of the potential barrier $U/\mu \sim 1$.
All the parameters used here are the same as Fig. \ref{fig3}(a).}
\label{fig6}
\end{figure}
\par
Now as one can see from Fig. \ref{fig6} when potential barriers are very weak ($U\ll \mu$) which leads to very transparent interfaces, CAR and the intensity of focusing is negligibly small. By increasing the potential strength the EHC and resulting CAR becomes more effective at the intermediate values $U\sim \mu$. Nevertheless for very strong barriers the competition between normal reflection and CAR is in favor of the first process. Therefore the intensity of the local hole current around  the focuses declines by further increase in the barriers strength and becomes fully suppressed for $U\gg \mu$. Unlike the hole current which reveals a maximum value around $U\sim \mu$, the electron local current, originated from EC, monotonically decreases with $U$ in particular in the regime of strong barriers 
($U > \mu$).
\subsection{On the experimental feasibility}
As it has been already mentioned, we show that a generic superconducting NISIN heterostructure, can focus both electrons and holes. Moreover, the focusing property of the junction is irrespective of the band structure and dispersion relation of the normal leads and the superconductor. In fact there is no need for special band structure or other features like the Dirac spectrum  considered in previous studies \cite{yeyati12}. Then devices based on conventional systems like 2DEG as well as graphene and other 2D materials can be used as the normal leads contacted to a superconducting region at the middle to see the lensing phenomena for electrons and holes.  
\par
Although the specific band structure is not an important requirement for the superconducting lens studied here, it should be noticed that this electronic lens could work properly under certain circumstances. Nevertheless, we will discuss that these requirements can be fulfilled  
in currently available experimental setups and subsequently our predictions can be checked experimentally. First of all we assume that the Fermi velocities inside normal and superconducting region are almost equal to each other. Fortunately this assumption can be essentially satisfied in the common superconducting heterostructures based on metals, semiconductors, and even structures based on graphene and topological insulators \cite{heersche,hasan14}. Another important requirement is the presence of effective potential barriers or equivalently insulating layers at the interfaces. As it has been shown in the previous part, the maximum feasibility of the focusing property occurs for $U\sim\mu$. However, in a practically wide range of barrier strengths, the focusing at least for the holes can be observed.
Very common an insulating layer is established due to the oxidation or mismatches in the electronic characters of the normal and superconducting layers. Therefore it seems that controlling the barrier strength would be challenging. However by changing the materials, different barrier strength can be achieved \cite{kleine09,fritz09} and even it is possible to have highly transparent interfaces like Al/Au \cite{zimansky06}. Therefore choosing suitable materials for the two normal and the superconducting leads, different barrier strengths for the two interfaces and particularly the extreme cases of NISN or NSIN structures with single barrier can be achieved. 
\begin{figure}[tb]
~~~\includegraphics[width=0.9\linewidth]{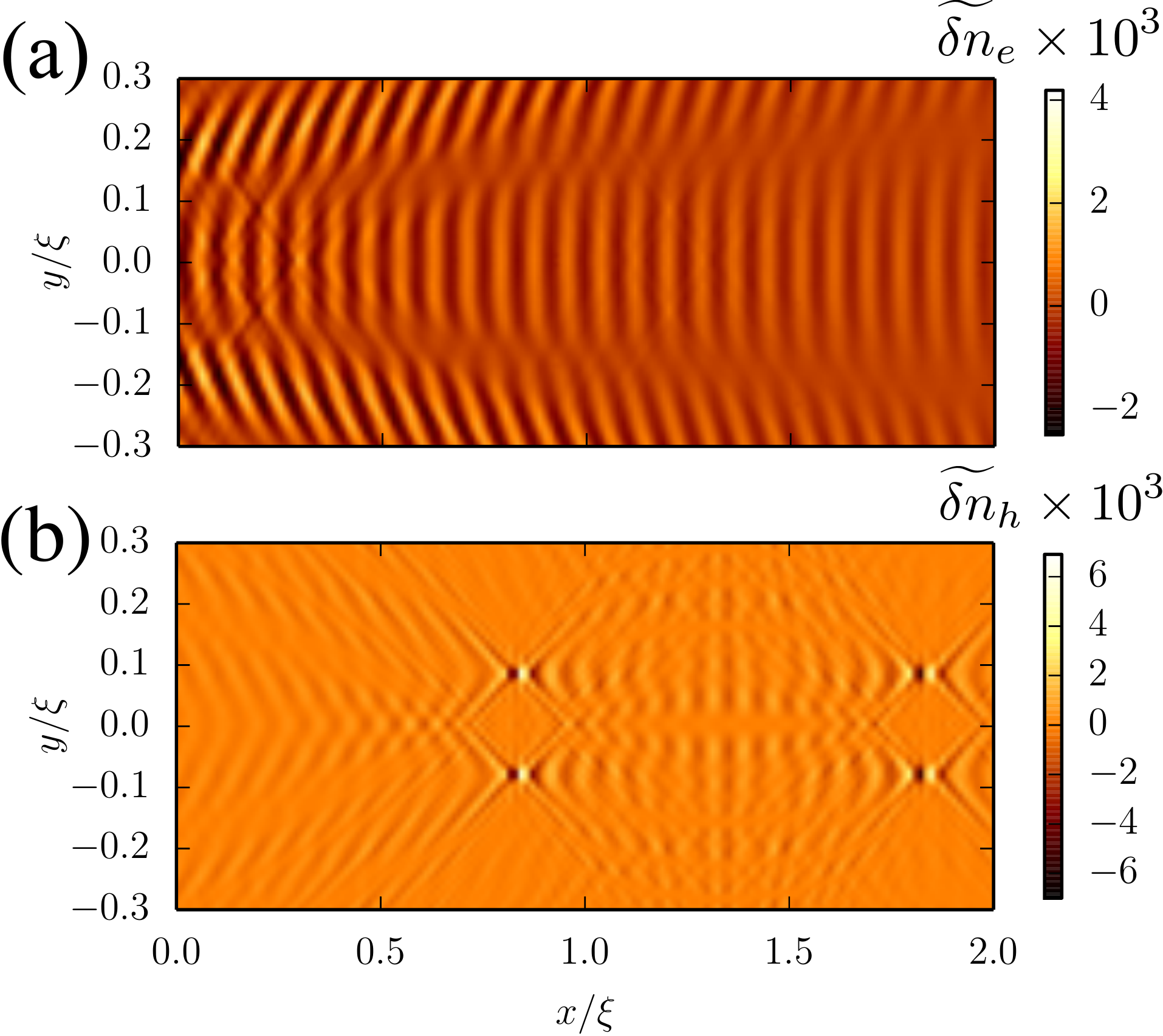}
\caption{(Color online) Spatial variations of (a) electron (b) hole densities inside the second normal contact N$_2$ when there are two injection points separated by $\delta y=5\lambda_F$ in y direction. All parameters are the same as those used for Fig. \ref{fig2}. The corresponding focuses of each injection can be clearly seen beside extra interferences of electron and hole waves induced by two point source.}
\label{fig7}
\end{figure}
\par
On top of the above mentioned requirements, there are some other parameters including disorder or finite size in $y$-direction which may affect the focusing character of the junction.
Although it has not been considered explicitly here, 
however similar to the graphene-based Veselago lenses, a finite disorder strength, suppresses the focusing signals and the focal points are washed out and broadened by the impurity scatterings \cite{yeyati12}. Hence it is crucial to have clean and almost impurity-free samples to see clear focusing patterns for holes and electrons. On the other hand in NISIN junctions with finite width in vertical direction, the back reflection of electronic waves from the boundaries could lead to further interferences. But as long as this vertical width is large enough compared to $W$, $x_0$, and $\xi$, the intensity of reflected waves will be much smaller than the focused waves coming from CAR. It is worth noting that the focuses positions remain unchanged at the presence of a finite vertical width. In fact as we can understand from the classical trajectories shown in Fig. \ref{fig1}, those with small enough angles will not see the boundaries before being focused by the S layer. All of these trajectories cross at the focal points and give rise to an intense focusing. However the trajectories with larger angles will be reflected by the horizontal boundaries and depending on their angles they will pass through different points. Therefore unless for very narrow junction in which most of the electronic waves are reflected, the finite vertical width cannot strongly influence the focusing pattern.
\par
Throughout the paper we have considered zero temperature which simplifies the Green's function calculations, while at finite temperatures one needs to calculate finite temperature Green's function which technically speaking needs to sum over Matsubara frequencies. But even without calculations it is possible to discuss over the effect of temperature on focusing, based on physical intuition. To this end we note that according to our model and the obtained results, the excitation energy can only slightly affect the CAR and EC signals. Moreover due to the electron-hole symmetry as a basic character of superconductors, for all values of the excitation energy inside the superconducting gap ($|\epsilon|<\Delta_0$), the positions of focuses do not depend on $\epsilon$. This can be simply understood from the semiclassical arguments and the fact that wave vectors dependence on $\epsilon$ is negligible. Now the main effect of temperature is to excite quasiparticles with energies below the thermal energy ($\epsilon\lesssim k_BT<\Delta_0$). Thus as long as the S layer is in its superconducting state, most of the thermally excited quasiparticles will have subgap energies and all of them will be focused almost in the same way. So we can say in brief that due to the electron-hole symmetry as a basic character of superconductors, the focusing properties of the junction cannot be influenced by thermal excitations when $T<T_c$ with $T_c$ indicating the critical temperature of the superconductor. As a result there will not be any observable broadening of the focal points due to the finite temperature. This is in fact an advantage of superconducting lens studied here in comparison with Veselago-like lenses in materials with chiral relativistic dispersion like graphene \cite{yeyati12,cheianov07,moghaddam10}. 
For instance in the \emph{pn} junctions based on graphene, even when one considers a symmetric potential profile, only electrons and holes at the Fermi level have the same magnitude momenta. Then in graphene-based Veselago lenses, the electron-hole symmetry is broken for the excitations above the Fermi level. Hence, depending on the excitation energy, the focusing takes place in slightly different points which leads to the broadening of the focuses at finite temperatures \cite{cheianov07}. 
\par
We finally comment on the possible experimental setups for the injection of electrons and detection of lensing effects. In order to inject electrons locally from a point source
one can either use a tip of scanning tunneling microscope or alternatively a small submicron size gate with a size comparable with the Fermi wavelength or below it. Even for larger electron sources, the focusing would take place but in a broadened form of the focuses inside N$_2$. Although we have not considered this case explicitly, however in order to understand the effect of a broad injection point, Fig. \ref{fig7} shows the density variations at the presence of two point sources separated by $\delta y=5 \lambda_F$ in vertical direction.
We see that irrespective of extra interferences, we will have well separated focuses corresponding to each of the injection points. Therefore we can simply expect similar imaging signature for more injection points and even a broad electron source. Similar to the injection itself, the spatial variation of the electron and hole local currents and densities inside N$_2$ can be detected via the scanning probe microscopes (SPMs) which have been already used for imaging electron motions in 2DEG \cite{topinka,aidala}.

\section{Conclusions}
\label{sec4}
In summary, we study the focusing of electron and hole excitations induced by CAR and multiple reflections from the interfaces of an NISIN junction when incoming electrons are locally injected inside the left metallic contact. It is shown that the barrier potentials at the interfaces, intensifies the focal points due to the CAR, while suppressing the EC signal. In fact, the barrier potentials provide the required abrupt changes in particle momentum during a CAR. In addition, the intensities and the configuration of electron and hole focuses  can be controlled by varying the location of electron injector, width of the superconductor, and the potential barriers. These findings suggest promising applications of the superconducting lenses for remote manipulation of the entanglement in spatially separated electron or hole pairs. 

\acknowledgments
We would like to acknowledge the encouragement and  instructive comments of late Malek Zareyan in the early stages of the work. H.C. and A.G.M. thank the International Center for Theoretical Physics (ICTP) for their hospitality and support during a visit in which part of this work was done. A.G.M. acknowledges financial support from Iran Science Elites Federation under Grant No. 11/66332

\bibliography{lens-cem}

\end{document}